\documentclass{ws-procs9x6}            
\usepackage{epstopdf}
\usepackage{graphicx}

\begin{document}
\title{The pseudoscalar meson and baryon octet interaction \\
in the unitary coupled-channel approximation}

\author{Bao-Xi Sun$^*$, Zheng-Ran Zhang, Hai-Lin Wu, Si-Yu Zhao and Fang-Yong Dong}

\address{College of Applied Sciences, Beijing University of
Technology,\\
 Beijing 100124, China\\
$^*$E-mail: sunbx@bjut.edu.cn}

\begin{abstract}
The pseudoscalar meson-baryon octet interaction is studied within a
nonlinear realized Lagrangian, and then the Bethe-Salpeter equation
is solved in the unitary coupled-channel approximation. In sector of
strangeness $S=-1$ and isospin $I=0$, only one pole is generated
dynamically in the 1400MeV region, which might correspond to the
$\Lambda(1405)$ particle. When the case of strangeness zero is
studied, the $s-$ and $u-$ potentials are taken into account in the
kernel, and a resonance state is produced in the 1500MeV region,
which might be a counterpart of the N(1535) particle.
\end{abstract}

\keywords{Chiral Lagrangian; Meson-baryon interaction; Hadronic
resonance; Bethe-Salpeter equation.}

\bodymatter

\section{Introduction}

The chiral perturbation theory has become a powerful tool to study
the pseudoscalar meson-baryon octet
interaction\cite{Weinberg,Scherer}. However, this theory is not
suitable to the energy region where the hadronic resonance appears.
Therefore, a non-perturbative resummation technique has to be taken
into account.
By solving the Bethe-Salpeter equation in the unitary
coupled-channel approximation, the hadronic resonance state can be
generated dynamically\cite{Kaiser,Oset97,Russian}. In the
calculation, the unitarity of the scattering amplitude is conserved,
which implies all terms in the expansion of the scattering amplitude
are included and no truncation is performed when the Bethe-Salpeter
equation is solved.

\section{The pole structure of $\Lambda(1405)$}

The pseudoscalar meson-baryon octet interaction with strangeness
$S=-1$ and isospin $I=0$ was studied in the unitary coupled-channel
approximation by solving the Bethe-Salpeter equation, and two poles
of the scattering amplitude were found on the complex energy plane
between the $\pi \Sigma$ and $\bar{K}N$ thresholds, which were
assumed to correspond to the $\Lambda(1405)$ particle.
Since 2016, a review article on the double-pole structure of the
$\Lambda(1405)$ particle had appeared in the Particle Data
Group(PDG) manual, where many theoretical research works on this
problem were cited\cite{PDG}.
However, In Ref.~\cite{DongSun}, the loop function of the
intermediate pseudoscalar meson-baryon octet is derived from the
dimensional regularization scheme strictly, and the relativistic
kinetic effect is taken into account when the Bethe-Salpeter
equation is solved. Finally, the revised loop function in the
dimensional regularization scheme takes the form of
\begin{equation}
\begin{aligned}
G_{l}={}&\frac{1}{32\pi^{2}\sqrt{s}}\left[(a_l+1)(m_{l}^{2}-M_{l}^{2})
+\left(m_{l}^{2}ln\frac{m_{l}^{2}}{\mu^{2}}
-M_{l}^{2}ln\frac{M_{l}^{2}}{\mu^{2}} \right)\right]\\&
+\left(\frac{s+M_{l}^{2}-m_{l}^{2}}{4M_{l}\sqrt{s}}
+\frac{1}{2}\right)G_{l}^{\prime},
\end{aligned}\label{eq:Our G result}
\end{equation}
where the $G_{l}^{\prime}$ is the original form of the loop
function\cite{OllerMeissner2001}, while $\sqrt{s}$ and $P$ are the
total energy and the momentum of the system in the center of mass
frame, $M_{l}$ and $m_{l}$ are the masses of the intermediate baryon
and pseudoscalar meson, respectively.
In Eq.~(\ref{eq:Our G result}), the subtraction constant $a_l$ and
the regularization scale $\mu$ can be regarded as one parameter in
the calculation.

By fitting the energy shift and width of the 1s state of kaonic
hydrogen measured precisely at the SIDDHARTA experiment at $DA\Phi
NE$\cite{SIDDHARTA} and the branch ratios at the $K^{-}p$
threshold\cite{Tovee,Nowak},
the subtraction constant can be determined, which are listed in
Table~\ref{table:subtraction}, while the regularization scale
$\mu=630$MeV is fixed.

\begin{table}
\tbl{The subtraction constants used in the calculation with the
regularization scalar $\mu=630$MeV.}
{\begin{tabular}{@{}c|cccccc@{}} \toprule
           &$\bar{K}N$ & $\pi \Lambda$ & $\pi \Sigma$ & $\eta
 \Lambda$  & $\eta \Sigma$ & $K \Xi$    \\
\colrule
 $a_l$    & -1.1  & -1.6 & -1.9 & -2.7 & -2.6 & -2.8    \\
\botrule
\end{tabular}}
\label{table:subtraction}
\end{table}

\begin{figure}[htbp]
\centering
\includegraphics[width=4.5in]{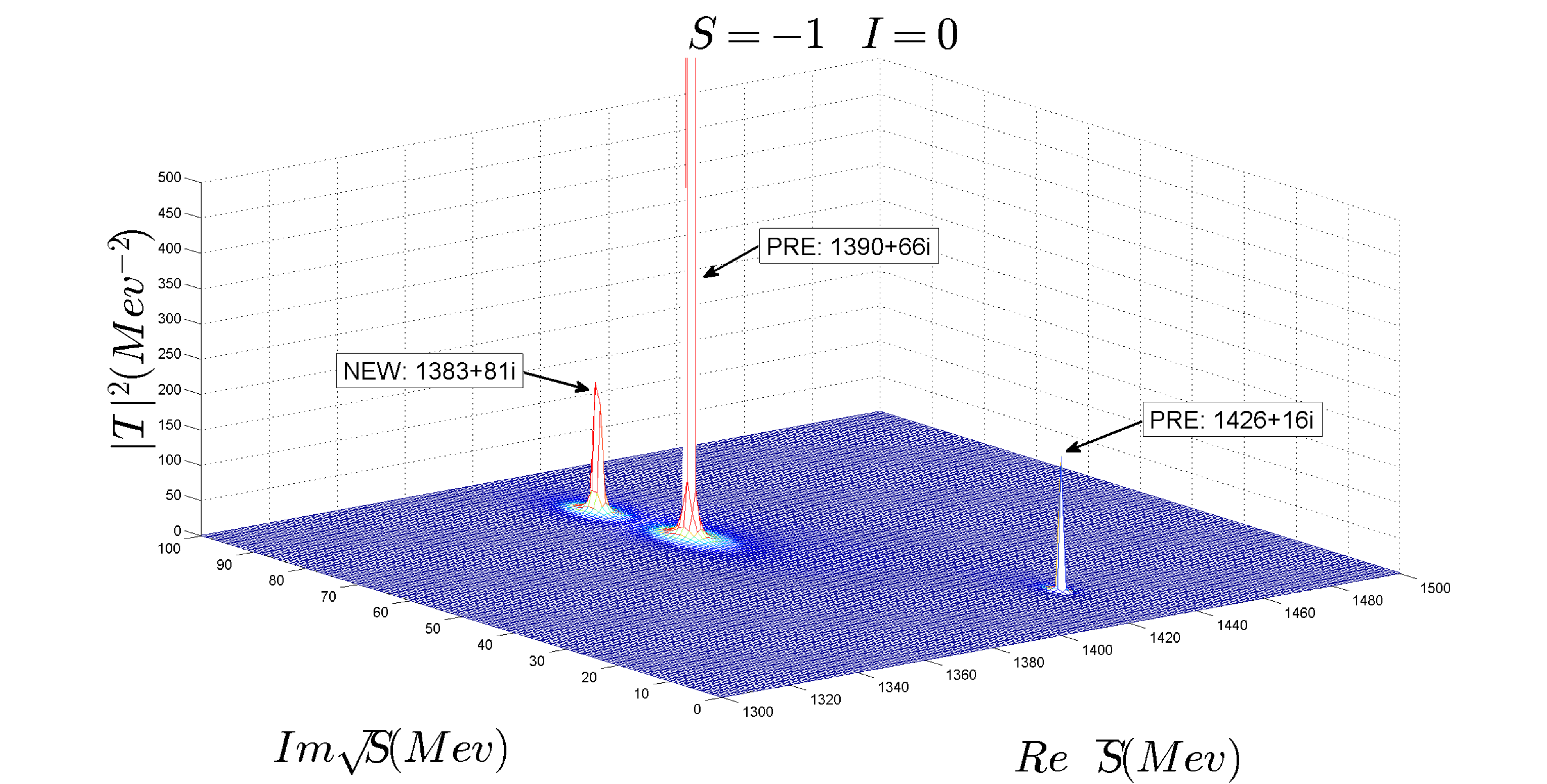}
\caption{Comparison of poles in the strangeness $S=-1$ and isospin
$I=0$ sector. The label $NEW$ represents the case calculated from
the loop function in Eq.~(\ref{eq:Our G result}), while the label
$PRE$ stands for the case of the original loop function $G^\prime_l$
with parameters in Ref.~\cite{OsetRamos2002}. } \label{aba:fig1}
\end{figure}

With the loop function in Eq.~(\ref{eq:Our G result}) and the
parameters in Table~\ref{table:subtraction}, only one resonance
state is generated dynamically between the $\pi \Sigma$ and
$\bar{K}N$ thresholds, which might be a counterpart of the
$\Lambda(1405)$ particle, as depicted in Fig.~\ref{aba:fig1}. It
implies that too many approximations had been made in the original
formula of the pseudoscalar meson-baryon octet loop function when
the Bethe-Salpeter equation is solved, and the double-pole structure
of the $\Lambda(1405)$ particle is model-dependent.

If one hopes the double pole structure of the $\Lambda(1405)$
particle is correct, the pseudoscalar meson-baryon octet interaction
must be studied in the unitary coupled-channel approximation by
solving the Bethe-Salpeter equation with the loop function where the
relativistic kinetic effect is taken into account, as used in
Ref.~\cite{DongSun}. Otherwise, the method without on-shell
approximation can also be used, i.e., this problem should be
recalculated by solving the integral formula of the Bethe-Salpeter
equation in the unitary coupled-channel approximation. If the
double-pole structure of the $\Lambda(1405)$ particle can be
reproduced by fitting the experimental data, our conclusion made in
Ref.~\cite{DongSun} can be certified to be incorrect.
In this sense, any attempt to certify the double-pole structure of
the $\Lambda(1405)$ particle with the original formula of the loop
function by solving the Bethe-Salpeter equation is unreasonable.

\section{$N(1535)$}

With the loop function in Eq.~(\ref{eq:Our G result}), we continue
to study the pseudoscalar meson-baryon octet interaction in the
strangeness $S=0$ sector\cite{SunZhaoWang}. In addition to the
Weinberg-Tomozawa term, the $s-$ channel and $u-$ channel potentials
are also taken into account in the calculation.

The $s-$ channel and $u-$ channel potentials are weaker than the
Weinberg-Tomozawa term in the pseudoscalar meson-baryon octet
interaction. However, if they are included as a kernel in addition
to the Weinberg-Tomozawa term when the Bethe-Salpeter equation is
solved in the unitary coupled-channel approximation, the subtraction
constants must be readjusted in order to produce the resonance state
at the reasonable position on the complex energy plane.

A resonance state is generated dynamically in the 1500MeV region,
which lies between the $\eta N$ and $K \Lambda$ thresholds and can
be regarded as a counterpart of the $N(1535)$ particle.
Since the $\pi N$ threshold is far lower than the energy region
where the N(1535) particle might be generated dynamically, it is
found that the pole position is not sensitive to the subtraction
constant $a_{\pi N}$.
Moreover, the mass of the $N(1535)$ particle is close to the $K
\Lambda$ threshold, and the subtraction constant $a_{K \Lambda}$ has
an important influence on the generation of the $N(1535)$ particle.
Therefore, the $a_{K \Lambda}$ value is stable in the calculation.

\section{Summary}

The interaction of the pseudoscalar meson and the baryon octet is
studied within a nonlinear realized Lagrangian. When the
Bethe-Salpeter equation is solved in the unitary coupled-channel
approximation, the relativistic kinetic correction is included in
the loop function of the intermediate meson and baryon, which
results in only one pole corresponding to the $\Lambda(1405)$
particle generated dynamically on the complex energy plane in the
strangeness $S=-1$ and isospin $I=0$ sector. When this method is
extended to study the case of strangeness zero, the $s-$ and $u-$
channel potentials are taken into account in addition to the
Weinberg-Tomozawa term. By adjusting the subtraction constants, a
resonance state is generated dynamically in the 1500MeV region,
which might be a counterpart of the N(1535) particle in the PDG
data.

\section*{Acknowledgments}
Bao-Xi Sun would like to thank Han-Qing Zheng for useful
discussions.

\end{document}